# Secure Channel for Molecular Communications

S. M. Riazul Islam
Dept. of Computer Engineering
Sejong University
Seoul, South Korea
riaz@sejong.ac.kr

Farman Ali
UWB Wireless Research Center
Inha University
Incheon, South Korea
farmankanju@gmail.com

Hyeonjoon Moon
Dept. of Computer Engineering
Sejong University
Seoul, South Korea
hmoon@sejong.edu

Kyung-Sup Kwak
UWB Wireless Research Center
Inha University
Incheon, South Korea
kskwak@inha.ac.kr

*Abstract*—Molecular communication in nanonetworks is an emerging communication paradigm that uses molecules as information carriers. Achieving a secure information exchange is one of the practical challenges that need to be considered to address the potential of molecular communications in nanonetworks. In this article, we have introduced secure channel into molecular communications to prevent eavesdropping. First, we propose a Diffie–Hellman algorithm-based method by which communicating nanomachines can exchange a secret key through molecular signaling. Then, we use this secret key to perform ciphering. Also, we present both the algorithm for secret key exchange and the secured molecular communication system. The proposed secured system is found effective in terms of energy consumption.

*Keywords— Nanonetworks, molecular communications, security, secure channel, eavesdropping.*

## I. INTRODUCTION

Molecular communication is promising to provide appropriate solutions for a wide range of applications including biomedical, industry, and environmental areas [1], [2]. Even, it can be integrated with the Internet of Things (IoT) to provide IoT-based healthcare services [3], [4] by implementing the concept of Internet of NanoThings [5]. This is an interdisciplinary research field and is significantly different from the electromagnetic (EM) communication system, since it utilizes molecules as carriers of information. However, molecular communication has a number of technical challenges to overcome. Introducing security into molecular communications is a fundamental challenge for researchers. There exist very few papers that focus on security aspects of this promising technology [6], [7]. These papers discuss several security and privacy issues and challenges in the context of molecular communications, in general. To the best of our knowledge, there exists no prominent work which attempts to mitigate any particular security risk.

In this article, we deal with eavesdropping, a special class of threats, that might exploit vulnerabilities to breach security in molecular communications. It is the act of secretly listening to an ongoing communication between two nanomachines. In EM communications, a secure channel is established to prevent eavesdropping. To do this, two communicating parties follow Diffie–Hellman key exchange protocol and generate a private key (a key only known to them). Then, information is encrypted using the key before transmission. However, in molecular communications, the main challenge is how to generate a private key so that the adversary cannot learn the exchanged key. Moreover, both key exchange process and encryption algorithms should be cost-effective in terms of computational complexity, and energy consumption. Authors in [8] make use of radio-frequency identification (RFID) noisy tags, similar to blocker tag suggested by authors in [9], to exchange secret key. These tags intentionally generate noise on the channel so that intruders in RFID communications can't understand the key. The same idea has been tailored to security requirements in near field communication (NFC) [10]. However, these works have been performed in the context of EM communications. In this article, we try to establish a secure channel in molecular communications to defend against eavesdropping.

## II. SYSTEM MODEL

We consider that the underlying system is time-slotted with specific slot duration and the participating nanomachines are in a stationary fluidic medium at any distance within the network coverage. We also consider that these nanomachines are perfectly synchronized and communicate with each other using same types of messenger molecules. Symbols are supposed to be transmitted upon on-off keying (OOK) modulation through the memoryless channel. In this scheme, information bit 1 is conveyed by liberating an impulse of $z_1$ number of molecules at the start of the slot, whereas no molecule is released for information 0. For the simplicity of presentation, we consider a full-duplex system; participating nanomachines can transmit and receive information at the same time. However, the proposed method, with a slight modification, can also equally be applied in a half-duplex system.

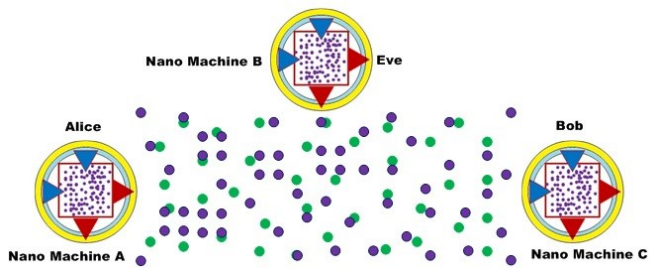

Fig. 1. Molecular communication with eavesdropping (A sends violet molecules, C-green molecules).

This work was supported in part by National Research Foundation of Korea-Grant funded by the Korean Government (Ministry of Science and ICT)-NRF-2017R1A2B2012337) and in part by Sejong University Faculty Research Fund.

In case of reception, a nanomachine counts the total number of messenger molecules received during the time slot. This received number of molecules, denoted by $z_2$, is then compared to z, a threshold number of molecules. If $z_2$ is less than z, the machine considers the received bit to be 0; otherwise, it decodes the received bit to be 1. We further assume that there exists at least a malicious nanomachine which uses a suitable detector to receive the transmitted molecules with the purpose of eavesdropping (see Fig. 1).

### III. EAVESDROPPING DISTANCE

This is very noticeable that eavesdropping is a significant matter, since molecular communication is a wireless approach in practice. A question that may arise is how a malicious nanomachine can decode the transmitted data out of received molecules. This can be achieved by two ways. First, the intruder can do required experiments prior to an attack. Second, the attacker can have prerequisite knowledge from literature investigation. Also, it's not unusual that the intruder will have the required detector to receive the molecules and the hardware/software arrangements to decode the received molecules, since this doesn't need any special kit.

Molecular communication is typically occurred between two nanomachines in near vicinity. A reasonable question is how close the malicious machine needs to be to be capable to detect the transmitted molecules. However, there is no exact answer to this question. The reason why we can't answer this question accurately is that there is a number of factors which determine the said distance. For instance, the distance might be influenced by the following factors, among others: molecular characteristics of the given transmitting nanomachine, number of molecules sent by transmitting nanomachine, detector characteristics of the intruder, quality of the intruder's nanomachine itself and location of the attack. Thus, any particular distance specified would only be usable for a definite set of the aforementioned parameters and cannot be utilized to develop common security strategies. Nevertheless, we can arguably say that eavesdropping can be performed up to a distance of more than a typical distance between two authentic nanomachines. In other words, a powerful attacker can still decode the information even when it operates from a relatively larger distance compared to intended receiving nanomachine..

### IV. PROPOSED SECURE CHANNEL

The idea is that both nanomachine A and nanomchine C transmit random data simultaneously. This is conceivable as nanomachines can launch and collect the molecules at the same time. In the setup phase, these two nanomachines synchronize on the exact timing of the bits and also on the energies (number of molecules) of the transmitted molecular signal. After the synchronization phase, machines A and C are able to send molecules at the same time with the same number of molecules. At the time of transmitting random bits of 0 (sending no molecules) or 1 (sending some predetermined number of molecules), both nanomachines also listen to molecular signal. Now, we consider all the possible cases below:

Case 1: When both machines transmit a zero molecular signal, the sum of these two molecular signals is zero and a malicious nanomachine, who is eavesdropping, would recognize that both nanomachines sent a zero. This case does not help the malicious machine B to understand that which machine is sending a bit of the secret key.

Case 2: When both machines transmit a one molecular signal, the sum of these two molecular signals is the double (two times the number of molecules set for sending a one) and the malicious nanomachine would recognize that both nanomachines sent a one. This case also does not help the malicious machine B to recognize that which machine is sending a bit of the secret key.

Case 3: An interesting case is happened when nanomachine A sends a one whereas nanomachine C sends a zero or when machine A sends a zero whereas machine C sends a one. In this situation, both nanomachines can find what the other machine has transmitted, since both machines are aware of what information they themselves have just sent. Conversely, the malicious node B only understands the sum of two molecular signals and it cannot factually find out which machine sent the one and which machine sent the zero.

This concept has been demonstrated in Fig. 2. While the top part of the figure shows the molecules released (violet color) by nanomachine A, whereas the middle part shows the molecules released (green color) by C. Machine A randomly transmits the eight bits: 1, 0, 1, 1, 0, 1, 0, and 1. Machine B randomly transmits the eight bits: 0, 1, 0, 1, 1, 1, 0, and 1. The bottom part of the figure displays the sum of the molecules released by both machines. This is the ultimate signal as observed by the malicious machine B. It clearly shows the resultant signal in case of A transmits 0, and C transmits 1 is the same as in the case of A transmits 1, and C transmits 0. Therefore, the malicious machine B can't differentiate between these two cases. Then, the two authentic machines now abandon all bits, where they transmitted the same number of molecules (corresponding to both A and C sent 0, and both A and C sent 1). However, both machines accept all bits, where they transmitted different number of molecules (corresponding to A sent 1 and C sent 0, and A sent 0 and C sent 1). They can consider either the bits transmitted by machine A or the bits transmitted by machine C as the secret key. This is basically a prior agreement as per security policies. In this fashion, both machines can exchange an encryption key of any desired length. Thus, in this example, transmitted bits, by the selected nanomachine, at bit indexes 0, 1, 2, and 4 constitutes the secret key, since the participatory machines sent different number of molecules at these bit indexes. Now, if a security scheme selects the nanomachine C for this purpose, the encryption key will be 0101. However, if the machine A is selected, the key is 1010. The key we have just obtained is 4-bits in length. Note that a secret key of any desired length can be achieved if both machines continue their operations until the target number of bits are stored. The flowchart in Fig. 3 shows the generalized algorithm of our proposed technique as described above. In this article, we will assume the desired key is 8 bits in length.

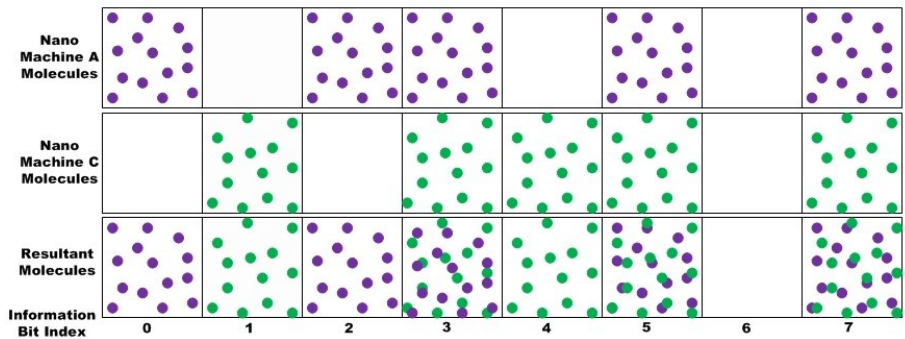

Fig. 2. The secret key exchange in molecular communication.

The next step after the private key generation is to encrypt the information bit to be transmitted before it goes onto the channel through molecular modulation. For this purpose, we use Exclusive OR (XOR) cipher, since it is simple to implement and XOR operation is inexpensive in terms of computation. The encrypted bits are obtained by using the following logical operations,

$$x_j = b_{kj} \oplus b_{ij} \text{ for } j = 0,1,\dots,7 \qquad (1)$$

where $x_j$, $b_{kj}$ and $b_{ij}$ are the $j$-th bit of the encrypted bits to be modulated for molecular transmission, the secret key, and information block of 8 bits, respectively. There will be many applications of molecular communications where stringent security will not be required. In those cases, a simple hiding operation is sufficient enough to hide the information from unauthorized parties. This in turn ensures that frequent changes of private key are not mandatory. Conversely, security-sensitive applications will require relatively frequent changes of secret keys. In the receiving side, decryption operation, presented in (2), will be performed on the information bits after molecular demodulation by using the same logical operations as applied in (1).

$$y_j = b_{kj} \oplus b_{ej} \text{ for } j = 0,1,\dots,7 \qquad (2)$$

where $y_j$, $b_{kj}$ and $b_{ej}$ are the j-th bit of the decrypted bits, the secret key, and demodulated information block of 8-bit, respectively. If there occurs error free communication, $y_j$ should be the same as $b_{ij}$. Our proposed secured molecular communication system thus eventually takes the form of Fig. 4. The serial-to-parallel (S/P) and parallel-to-serial (P/S) converters have been used, since the hardware XOR cipher performs 8-bit parallel XOR operations. However, this is a design issue. The same XOR cipher can alternatively be accomplished by using a single XOR gate. It is worth mentioning that the processing time to place secured molecular information onto the channel becomes negligible compared to the usual baseband information processing time, since there is no complicated computation in ciphering operations and parallel-to-serial and serial-to-parallel operations also occur instantly. Moreover, the use of hardware ciphering instead of software counterpart further reduces the associated time. Thus, our secured molecular communication system is effective in terms of information processing time.

## V. ENERGY CONSUMPTION ANALYSIS

Let $E_b^T$ and $E_b^C$ be the energy required to transmit one bit of information, and energy required to compute one bit of information, respectively. Also, $N$ denotes the total number of information bits to be transmitted and the key length is designated by $K$. And the number of key generation to complete the information transmission is $M$. Since information bits (1's or 0's) are randomly generated on an equiprobable basis and transmissions of two out of four possible cases (during key exchange) are discarded, participating nodes should exchange $2n$ bits of information, on an average, to share $n$ bits of key. Therefore, the required energy to exchange the desired key once is $E_K = 2 \times K \times E_b^T$. Ciphering and deciphering $N$ information bits by using XOR logical operations require $E_C = 2 \times N \times E_b^C$ energy. Thus, the total energy required to transmit the information with security becomes

$$E_T^S = \text{Energy for Information Transmission}$$
$$+ M \times E_K + E_C$$
$$= N \times E_b^T + M \times (2 \times K \times E_b^T) + 2 \times N \times E_b^C. \quad (3)$$

Considering the fact that the energy required to transmit one bit of information is analogous to the energy required to carry 1000 logical operations [11], we use $E_b^C = 0.001 \times E_b^T$ in (3) and get

$$E_T^S = (1.002N + 2KM)E_b^T \qquad (4)$$

In case of no security, the total energy required to transmit the information is

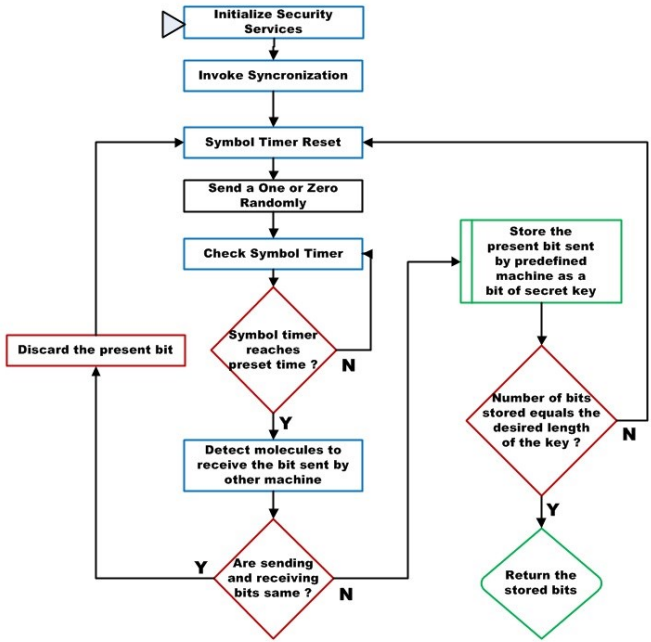

Fig. 3. The algorithm for secret key exchange.

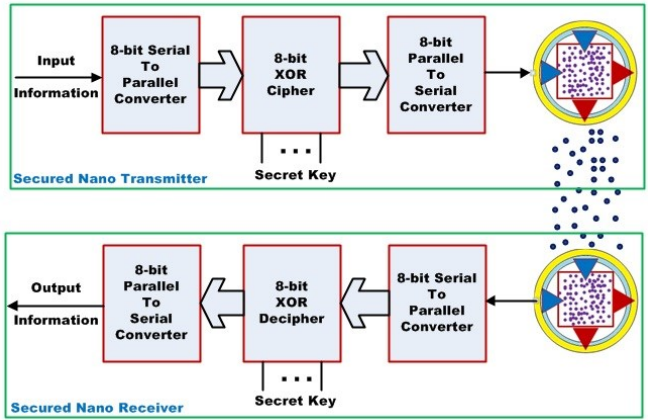

Fig. 4. The molecular communication system with the secure channel.

$E_T^0 =$ Energy for Information Transmission $= N \times E_b^T$ (5)

## VI. Performance Evaluation

To assess the proposed method, we perform computer simulation. We consider same types of molecules of the same size with OOK modulation scheme in a 2-dimensional confined space and 125 molecules/bit ($E_b^T$) is used on an average. The sizes of the messenger molecules are anticipated to be analogous to that of the fluid molecules. The threshold number of molecules is supposed to be $z = 20$. Transmitter and receiver are synchronized [12] so that the starting time of each molecular symbol at each communicating node becomes the same and the track of secret key sharing is maintained. The 4K information bits, divided into 4 frames, are transmitted. The frequent change of secret key is implemented by generating new keys after each 2 frames.

The Fig. 5 presents the total energy requirements for secured molecular communication system under varying key length. It clearly shows that no additional energy is required to transmit secured molecular information in the proposed secure channel fashion. This is because the number of bits to send the same amount information remains unchanged. However, the secured molecular communication system needs energy in exchanging the secret keys and in processing the information for encryption. In case of simple information hide operation, where the key should be exchanged only once, prior to information transmission, the same key can relatively be used for a long time. As a result, the amount of additional energy is negligible compared to the total energy requirement for information transmission. Even, the amount of extra energy consumption is still very low for the cases where there might be a provision of frequent changes of shared keys (different key after a few frames), since the key is very short in length compared to the total length of several frames of interest. Thus, the proposed method is found energy efficient. Both simulated and analytical results are well matched at every case.

## VII. Concluding Remarks

In this article, we have proposed a secured molecular communication system to defend against eavesdropping. The participating nanomachines exchange a secret key through molecular signaling in such a way that adversary cannot understand the key. This key exchange mechanism doesn't require significant additional energy. Also, the use of XOR ciphering to encrypt and decrypt the data using the generated secret key make the system simple and effective in terms of energy consumption. The proposed system can effectively be used in molecular communication systems where simple hide operations are sufficient or more stringent security are required.

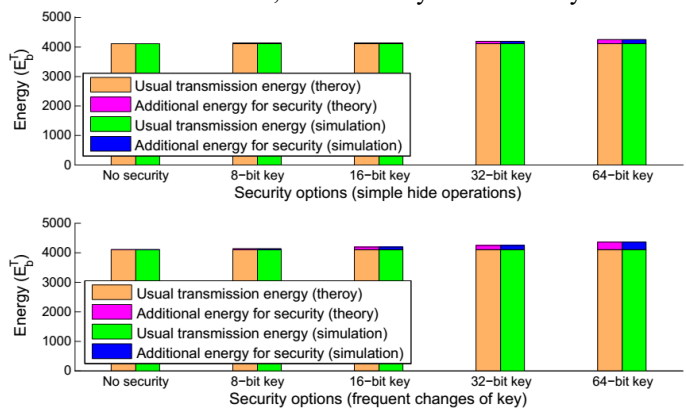

Fig. 5. Comparison of energy consumption.